\documentclass[10pt]{article}
\usepackage{amsmath}
\usepackage{amsfonts}
\usepackage{amssymb}
\usepackage[dvips]{epsfig}
\usepackage[dvips]{graphics}
\usepackage{graphicx}

\begin{document}

\title{Electromagnetic field at Finite Temperature: A first order approach}
\author{R. Casana\thanks{%
casana@ift.unesp.br}~, B.M. Pimentel\thanks{%
pimentel@ift.unesp.br}~ and J.S. Valverde\thanks{%
valverde@ift.unesp.br} \\
\textit{{\small Instituto de F\'{\i}sica Te\'orica (IFT/UNESP),
UNESP - S\~ao Paulo State University}} \\
\textit{\small Rua Pamplona 145, CEP 01405-900, S\~ao Paulo, SP, Brazil}}
\date{}
\maketitle

\begin{abstract}
In this work we study the electromagnetic field at Finite Temperature via
the massless DKP formalism. The constraint analysis is performed and the
partition function for the theory is constructed and computed. When it is
specialized to the spin 1 sector we obtain the well-known result for the
thermodynamic equilibrium of the electromagnetic field.
\end{abstract}

\section{Introduction}

Quantum Field Theory at Finite Temperature was motivated by the increasing
interest in studying the properties of matter under extreme conditions as,
for example, at very high temperature or density. The pioneering works
joining together the Statistical and Quantum Field Theory were developed
mainly by Matsubara \cite{Matsubara} in a non relativistic context and, the
relativistic case by Fradkin \cite{Fradkin} who, via the functional
approach, studied the different methods for calculating the Thermal Green's
functions as well as the structure of the Ward identities in QED$_{4} $.
Later other works within the Thermal Field Theory appeared \cite{Weinberg}
whose principal interest was to explore the possibility of restoring some
broken symmetries that occur at zero temperature, as for example the $%
SU(2)\times U(1)$ symmetry of weak interaction. The Finite Temperature gauge
theories and the problems concerning to the choice of a physical gauge and
its dependence was analyzed by Bernard \cite{Bernard}, in particular, the
free electromagnetic field.

On the other hand, at zero temperature, there is an alternative way to study
the properties for the electromagnetic field which is known as the massless
Duffin-Kemmer-Petiau (DKP) theory \cite{Chandra} that is not a trivial limit
of the massive Duffin-Kemmer-Petiau theory (DKP) \cite{Petiau} that appears
as an alternative formalism for the description of the spin 0 and spin 1
particles in a unified formulation. The DKP theory gives a first order
linear equation and it is very similar to the Dirac one but the $\beta^{\mu}$
matrices satisfy a different algebraic relation. The massive and massless
case of the theory were considered in the works \cite%
{minkowski,riemann,R-cartan} where the equivalence of the DKP theory with
the theories like Klein-Gordon-Fock (KGF) and Maxwell was proved in
Minkokwski space-time \cite{minkowski} and studied in curved space-time such
as Riemann \cite{riemann} and Riemann-Cartan \cite{R-cartan}.

At Finite Temperature the massive case is treated in the work \cite{Casana}
where the Bose-Einstein condensation is investigated in the spin 0 sector.
The equivalence of many-photons Thermal Green's functions of the DKP and KGF
theories was also proved for the scalar sector \cite{Fainberg1} calculating
the polarization operator at 1-loop order and, in \cite{last} is shown the
equivalence of many-gluons Green's functions in the DKP and KGF Statistical
Quantum Field Theories

As above mentioned, all accomplished studies on the massless DKP theory were
made at zero temperature. The aim of this work is to study the
thermodynamics of the electromagnetic field by using the massless DKP
theory. The paper is organized as following: In section 2, we give a brief
review of the massless DKP formalism considering a theory with only one real
DKP field. In section 3, the constraint analysis is performed for the DKP
theory and it has been shown that the model has two first class constraints.
In section 4, we calculate the partition function and, finally, we give our
conclusions and commentaries.

\section{The Massless DKP Theory}

The massless DKP theory is described in Minkowski space-time by the
following Lagrangian density \cite{Chandra}
\begin{equation}
\mathcal{L}_{M}=i~\bar{\psi}~\ \gamma \beta ^{\mu }\partial _{\mu }\psi
-i\partial _{\mu }\bar{\psi}\beta ^{\mu }\gamma \psi -\bar{\psi}\gamma \psi
~\ \ ,  \label{d1}
\end{equation}%
where $\bar{\psi}=\psi ^{\dag }\eta _{0}\,$ with $\,\eta _{0}=\left( 2\beta
_{0}^{2}-1\right) $. The $\beta ^{\mu }$ and $\gamma $ are singular square
matrices satisfying the following algebra
\begin{equation}
\beta ^{\mu }\beta ^{\lambda }\beta ^{\nu }+\beta ^{\nu }\beta ^{\lambda
}\beta ^{\mu }=\eta ^{\mu \lambda }\beta ^{\nu }+\eta ^{\nu \lambda }\beta
^{\mu }~\ \ ,  \label{d2}
\end{equation}%
\begin{equation}
\beta ^{\mu }\gamma +\gamma \beta ^{\mu }=\beta ^{\mu }~\ \ ,\quad \gamma
^{2}=\gamma ~\ \ .  \label{d3}
\end{equation}%
Due to the singular character of the $\beta ^{\mu }$ matrices the transition
of the massive case to the massless theory is non trivial and demands a
different treatment. On the other hand the representations of the DKP
algebra are reducible and contain the sectors of spin 0 and spin 1 in their
structure.

From the Lagrangian (\ref{d1}) we obtain the equation of motion for the
massless DKP field
\begin{equation}
i\beta ^{\mu }\partial _{\mu }\psi -\gamma \psi =0~\ \ .  \label{d5}
\end{equation}%
It can be shown that the Lagrangian (\ref{d1}) and the massless DKP equation
remain invariants under the following gauge transformation
\begin{equation}
\psi \rightarrow \psi ^{\prime }=\psi +\left( 1-\gamma \right) \Phi ~\ \ ,
\label{d6}
\end{equation}%
iff the field $\Phi $ satisfies the condition
\begin{equation}
i\beta ^{\mu }\partial _{\mu }\left( 1-\gamma \right) \Phi =0~\ \ .
\label{d7}
\end{equation}

When the fields under consideration are no charged we have a real DKP field{%
\footnote{{When the DKP field is real, the $\beta ^{\mu }$ matrices must
been satisfied $\beta _{0}^{T}=-\beta _{0}\,,\;\,\beta _{k}^{T}=\beta _{k}\,$%
.\thinspace\ The representation of the $\beta ^{\mu }$ matrices for
the spin 1 sector is given in the appendix A, where we also
included the $\gamma $ matrix.}}} $\psi $, in such situation the
Lagrangian (\ref{d1}) takes the following form
\begin{equation}
\mathcal{L}=\frac{i}{2}\psi ^{T}\left( \eta ^{0}\gamma \beta ^{\mu }\right)
\partial _{\mu }\psi -\frac{i}{2}\partial _{\mu }\psi ^{T}\left( \eta
^{0}\beta ^{\mu }\gamma \right) \psi -\frac{1}{2}\psi ^{T}\left( \eta
^{0}\gamma \right) \psi  \label{d8}
\end{equation}

\section{Constraint Analysis}

We proceed the study of the constraint analysis to the real massless DKP
theory from the Lagrangian (\ref{d8}) which is written as
\begin{equation}
\mathcal{L}=\frac{i}{2}\psi ^{a}\left( \eta ^{0}\gamma \beta ^{\mu }\right)
_{ab}\partial _{\mu }\psi ^{b}-\frac{i}{2}\partial _{\mu }\psi ^{a}\left(
\eta ^{0}\beta ^{\mu }\gamma \right) _{ab}\psi ^{b}-\frac{1}{2}\psi
^{a}\left( \eta ^{0}\gamma \right) _{ab}\psi ^{b}~\ \ .  \label{c1}
\end{equation}%
as usual we define the canonical momentum $\pi _{a}$ as
\begin{equation}
\pi _{a}=\frac{\delta L}{\delta \dot{\psi}^{a}}=-i\left( \beta ^{0}\gamma
\right) _{ab}\psi ^{b}~\ \ ,  \label{c2}
\end{equation}%
from which a set of primary constraints appear $\theta $%
\begin{equation}
\theta _{a}=\pi _{a}+i\left( \beta ^{0}\gamma \right) _{ab}\psi ^{b}~\ \ ,
\label{c3}
\end{equation}%
because the two different representations for the $\beta ^{\mu }$ matrices
we have for the spin $0$ sector that $a=\varphi ,0,1,2,3$ and for spin $1$
sector that $a=$ $0,1,2,...,9$.

The canonical Hamiltonian density $\mathcal{H}_{C}$ that follows from the
Lagrangian (\ref{c1}) is given by
\begin{equation}
\mathcal{H}_{C}=\frac{i}{2}\partial _{k}\psi ^{a}\left( \eta ^{0}\beta
^{k}\gamma \right) _{ab}\psi ^{b}-\frac{i}{2}\psi ^{a}\left( \eta ^{0}\gamma
\beta ^{k}\right) _{ab}\partial _{k}\psi ^{b}+\frac{1}{2}\psi ^{a}\left(
\eta ^{0}\gamma \right) _{ab}\psi ^{b}  \label{c4}
\end{equation}%
and considering the set of constraints (\ref{c3}) we have the primary
Hamiltonian density $\mathcal{H}_{P}\ $as
\begin{equation}
\mathcal{H}_{P}=\mathcal{H}_{C}+\lambda ^{a}\theta _{a}~\ \ ,  \label{c5}
\end{equation}%
where $\lambda ^{a}$ are the Lagrange multiplier.

The Poisson bracket (PB) for the primary constraints results in
\begin{equation}
\left\{ \theta _{a}\left( \mathbf{x}\right) ,\theta _{b}\left( \mathbf{y}%
\right) \right\} =i(\beta ^{0})_{ab}\delta \left( \mathbf{x-y}\right) ~\ \ .
\label{c6}
\end{equation}%
To investigate the possibility of obtaining more constraints in the theory
we apply the preservation in time of $\theta _{a},$ i.e.
\begin{equation}
\dot{\theta}_{a}\left( \mathbf{x}\right) =\left\{ \theta _{a}\left( \mathbf{x%
}\right) ,H_{P}\right\} ~\ \ ,
\end{equation}%
where $H_{P}=\displaystyle{\int \!\!d^{3}\mathbf{z}\,\mathcal{H}_{P}}$ is
the primary Hamiltonian. Thus, the stability condition provide
\begin{equation}
\dot{\theta}_{a}=i\left( \eta ^{0}\beta ^{k}\right) _{ab}\partial _{k}\psi
^{b}-\left( \eta ^{0}\gamma \right) _{ab}\psi ^{b}+i\beta _{ab}^{0}\lambda
^{b}\approx 0~\ \ ,  \label{c8}
\end{equation}%
being $\beta ^{\mu }$ singular matrices we conclude that not all $\lambda $\
coefficients can be obtained from the relation (\ref{c8}) and more
constraints appear. These new constraints are selected by means of the
projector $\mathbf{M}=\left( 1-\beta _{0}^{2}\right) $, thus, we obtain
\begin{equation}
\theta _{a}^{(2)}=\left( \mathbf{M}\beta ^{k}\right) _{ac}i\partial _{k}\psi
^{c}-\left( \mathbf{M}\gamma \right) _{ac}\psi ^{c}  \label{c9}
\end{equation}%
that are a set of secondary constrains. When the preservation in time of
these secondary constraints is imposed no more constraints appear in the
theory.

Now we calculate the PB for all primary and secondary constraints
\begin{eqnarray}
\left\{ \theta _{a}\left( \mathbf{x}\right) ,\theta _{b}^{(2)}\left( \mathbf{%
y}\right) \right\} &=&\left[ \left( i\beta ^{k}\partial _{k}^{x}+\gamma
\right) \mathbf{M}\right] _{ab}\delta \left( \mathbf{x}-\mathbf{y}\right) ~\
\ ,  \notag \\[0.3cm]
\left\{ \theta _{b}^{(2)}\left( \mathbf{x}\right) ,\theta _{a}\left( \mathbf{%
y}\right) \right\} &=&\left[ \mathbf{M}\left( i\beta ^{k}\partial
_{k}^{x}-\gamma \right) \right] _{ab}\delta \left( \mathbf{x}-\mathbf{y}%
\right) ~\ \ ,  \label{c10} \\[0.3cm]
\left\{ \theta _{a}^{(2)}\left( \mathbf{x}\right) ,\theta _{b}^{(2)}\left(
\mathbf{y}\right) \right\} &=&0~\ \ .  \notag
\end{eqnarray}%
We can write all set of constraints as $\zeta _{a}=\left\{ \theta
_{a},\theta _{a}^{(2)}\right\} $ such that its matrix is
\begin{equation}
\{\zeta _{a}\left( \mathbf{x}\right) ,\zeta _{b}\left( \mathbf{y}\right) \}=%
\left[
\begin{array}{ccc}
i\,\beta ^{0} & \, & (i\beta ^{k}\partial _{k}^{x}+\gamma )\mathbf{M} \\
& \, &  \\
\mathbf{M}(i\beta ^{k}\partial _{k}^{x}-\gamma ) & \, & 0%
\end{array}%
\right] \delta (\mathbf{x-y})~\ \   \label{cc-1}
\end{equation}%
and the determinant of $\{\zeta _{a},\zeta _{b}\}$ is zero. But the
constraint structure depends of the chosen representation for the
DKP algebra as we will see go on.

\subsection{Spin 1 sector{\protect\footnote{%
For the spin 0, the representation for the $\beta^\mu$ matrices is $5\times{5%
}$. The rank of the constraint matrix $\{\zeta_a,\zeta_b\}$ is 8, the null
space has two trivial constraints which are irrelevant, and the set of
constraints is second class.}}}

For this sector the representation of the $\beta^\mu$ matrices is $10\times{%
10}$ and the spin 1 DKP field is a column matrix with ten real components
\begin{equation}
\psi =\left( \psi ^{0},\psi ^{1},\psi ^{2},\psi ^{3},\psi ^{4},\psi
^{5},\psi ^{6},\psi ^{7},\psi ^{8},\psi ^{9}\right) ^{T}~\ \ .
\end{equation}

Consequently, in this sector, from (\ref{c3}) we observe that there are ten
primary constraints and write explicitly
\begin{eqnarray}
&\theta _{0}=\pi _{0}~\ \ ,&  \label{e1-1} \\[0.3cm]
&\theta _{1}=\pi _{1}+\psi ^{7}~\ \ ,\quad \theta _{2}=\pi _{2}+\psi ^{8}~\
\ ,\quad \theta _{3}=\pi _{3}+\psi ^{9}~\ \ \ ,&  \label{e1-2} \\[0.3cm]
&\theta _{n}=\pi _{n}~\ \ ,\quad n=4,5,6,7,8,9~\ \ .&  \label{e1-3}
\end{eqnarray}

And from (\ref{c9}) we obtain four secondary constraints given by
\begin{eqnarray}
&\theta _{0}^{(2)}=-\partial _{1}\psi ^{7}-\partial _{2}\psi ^{8}-\partial
_{3}\psi ^{9}~\ \ ,&  \label{e2-1} \\[0.3cm]
&\theta _{4}^{(2)}=\partial _{3}\psi ^{2}-\partial _{2}\psi ^{3}-\psi ^{4}\
\ ,\quad ~\theta _{5}^{(2)}=\partial _{1}\psi ^{3}-\partial _{3}\psi
^{1}-\psi ^{5}\ \ ,\quad \theta _{6}^{(2)}=\partial _{2}\psi ^{1}-\partial
_{1}\psi ^{2}-\psi ^{6}~\ \ ,~\quad \quad &  \label{e2-2}
\end{eqnarray}

To classify these constraints as first and second class we perform
the calculation of the PB between all these primary and secondary
constraints such as it is shown by the matrix $\{\zeta _{a}\left(
\mathbf{x}\right) ,\zeta _{b}\left( \mathbf{y}\right) \}$ in
(\ref{cc-1}). In our case, the rank of the matrix $\{\zeta
_{a},\zeta _{b}\}$ is 12 which is the number of second class
constraints and, the dimension of the null space is 2 that gives the
number of the first class constraints. The null space is formed by
the constraint $\theta _{0}=\pi _{0} $ and by the linear combination
of second class constraints $\partial _{1}\theta _{1}+\partial
_{2}\theta _{2}+\partial _{3}\theta _{3}+\theta _{0}^{\left( 2\right) }~$ that defines $~$%
another first class constraint $G=\partial _{k}\pi_{k}$. Thus, we obtain two
first class constraints
\begin{equation}
\theta _{0}=\pi _{0}~\ \ \ ,\quad G=\partial _{k}\pi_{k}~\ \ , \; k=1,2,3
\label{e3}
\end{equation}%
and twelve second class constraints given by the equations (\ref{e1-2}), (%
\ref{e1-3}) and (\ref{e2-2}).

The projectors of the spin 1 sector are defined as
\begin{eqnarray}
R^{\mu } &=&\left( \beta ^{1}\right) ^{2}\left( \beta ^{2}\right) ^{2}\left(
\beta ^{3}\right) ^{2}\left[ \beta ^{\mu }\beta ^{0}-\eta ^{\mu 0}\right]
\notag \\
&& \\[-0.2cm]
R^{\mu \nu } &=&R^{\mu }\beta ^{\nu } \qquad,\;\qquad\mu,\nu=0,1,2,3.  \notag
\end{eqnarray}
such that the field $\psi^\mu=R^{\mu }\psi$ is a Lorentz vector and $%
\psi^{\mu\nu}=R^{\mu \nu }\psi $ is an antisymmetric second-rank
Lorentz tensor; when multiplied with the $\gamma$ matrix we also
get $R^{\mu }\gamma\psi=0$ and $R^{\mu \nu }\gamma\psi=R^{\mu \nu
}\psi$.

Using the projectors $R_{\mu }$ and $R_{\mu \nu }$ and, from the relation (%
\ref{d6}) we conclude that only the vector components of the DKP field are
transformed as it is shown to follow
\begin{eqnarray}
\psi _{\mu }^{\prime } &=&\psi _{\mu }+\Phi _{\mu }~\ \ ,\quad   \notag \\
&& \\[-0.2cm]
\psi _{n}^{\prime } &=&\psi _{n}\quad,\qquad n=4,5,6,7,8,9\quad ,
\notag
\end{eqnarray}%
and from (\ref{d7}) we get $R_{\mu }\Phi =\Phi _{\mu }=\pm \partial _{\mu
}\Lambda ,$\ being $\,\Lambda \,$ an arbitrary scalar function, thus, we can
conclude that the theory under consideration is a local $U\left( 1\right) $
gauge field theory. Then, we impose the following gauge fixing conditions
\begin{equation}
\Omega _{1}=\partial _{k}\psi ^{k}\qquad ,\qquad \;\Omega _{2}=\psi
^{0}\quad ,  \label{gfcs}
\end{equation}%
such that the set $\chi _{A^{\prime }}=\left\{ \theta _{0},G,\Omega
_{1},\Omega _{2}\right\} $ is second class, thus, the PB matrix of the set
is given by
\begin{equation}
D_{A^{\prime }B^{\prime }}=\left\{ \chi _{A^{\prime }}\left( \mathbf{x}%
\right) ,\chi _{A^{\prime }}\left( \mathbf{y}\right) \right\} =\left[
\begin{array}{cccc}
0 & 0 & 0 & -1 \\
0 & 0 & -\bigtriangleup _{x} & 0 \\
0 & \bigtriangleup _{x} & 0 & 0 \\
1 & 0 & 0 & 0%
\end{array}%
\right] \delta \left( \mathbf{x-y}\right)   \label{det-1}
\end{equation}%
where $\bigtriangleup =\left( \partial _{k}\right) ^{2}=\left(
\partial _{1}\right) ^{2}+\left( \partial _{2}\right) ^{2}+\left(
\partial _{3}\right) ^{2}$, computing the functional determinant
we get
\begin{equation}
\det D_{A^{\prime }B^{\prime }}=\left[ \det \bigtriangleup \right] ^{2}~\ \ .
\label{det-2}
\end{equation}

\section{The Partition Function}

Now we study the thermodynamic equilibrium of the electromagnetic field
using the DKP formalism. Such as we see the constraint analysis of the
theory gives for the spin 1 sector two first class constraints
characterizing a local $U(1)$ gauge theory. Then, we write the partition
function for the massless real DKP field using the Hamiltonian formalism
\begin{equation}
Z=\!\!\!\!\!\int\limits_{periodic}\!\!\!\!\!\mathcal{D}\psi \mathcal{D}\pi
\delta \left( \Theta _{A}\right) \delta \left( \chi _{A^{\prime }}\right)
\left( \det C_{AB}\right) ^{1/2}\left( \det D_{A^{\prime }B^{\prime
}}\right) ^{1/2}\exp \left\{ \int_{\beta }d^{4}x~\left( i\pi _{a}\partial
_{\tau }\psi ^{a}-\mathcal{H}_{C}\right) \right\} ~\ \ ,  \label{pa1}
\end{equation}%
where $\mathcal{H}_{C}$ is given by (\ref{c4}) and
\begin{equation}
\mathcal{D}\psi =\mathcal{D}\psi ^{a},\quad \mathcal{D}\pi =\mathcal{D}\pi
_{a}\quad ,\quad a=0,1,2,...,9~\ \ .  \label{pa2}
\end{equation}%
The fields $\psi $ are restricted by the periodicity condition
\begin{equation}
\psi \left( 0,\mathbf{x}\right) =\psi \left( \beta ,\mathbf{x}\right) ~\ \ ,
\label{pa3}
\end{equation}%
where $\Theta _{A}=\left\{ \theta _{a},\theta _{b}^{(2)}\right\} $ is the
set of second class constraints given by the equations (\ref{e1-2}), (\ref%
{e1-3}) and (\ref{e2-2}); the set $\chi _{A^{\prime }}=\left\{
\theta _{0},G,\Omega _{1},\Omega _{2}\right\} $ is given by the set
of first class constraints (\ref{e3}) and its respective gauge
fixing conditions (\ref{gfcs}). The matrix $C_{AB}=\left\{ \Theta
_{A},\Theta _{B}\right\} $ can be obtained from (\ref{cc-1}) and its
determinant is $\det C_{AB}=1$ . The matrix $D_{A^{\prime }B^{\prime
}}$ $=\left\{ \chi _{A^{\prime }},\chi _{B^{\prime }}\right\} $ and
its determinant $\det D_{A^{\prime }B^{\prime }} $ are given by the
equations (\ref{det-1}) and (\ref{det-2}), respectively.

But, it is interesting to perform the calculation of the partition
function in a manifest covariant way. Thus, it is possible to show
that the equation (\ref{pa1}) becomes
\begin{equation}
Z=N\left( \beta \right) \!\!\!\!\!\int\limits_{periodic}\!\!\!\!\!\mathcal{D}%
\psi \mathcal{\delta }\left( F\left[ \psi ^{A}\right] \right) \det
\left\vert \frac{\partial F^{g}}{\partial \Lambda }\right\vert \exp \left\{
\int_{\beta }d^{4}x\left[ \frac{1}{2}\psi ^{T}\eta ^{0}\left( i\beta
^{A}\partial _{A}-\gamma \right) \psi \right] \right\} ~\ \ ,  \label{p14}
\end{equation}%
where $F\left[ \psi ^{A}\right] $ is an arbitrary gauge fixing condition.
Here we consider
\begin{equation}
F\left[ \psi ^{A}\right] =\frac{1}{\sqrt{\zeta }}\partial _{A}\psi ^{A}-f~\
\ \ \ \ \ ,~\ \ \ \ \ \ \ F^{g}\left[ \psi ^{A}\right] =F\left[ \psi ^{A}%
\right] -\frac{1}{\sqrt{\zeta }}\square \Lambda \;~~.  \label{p15}
\end{equation}%
with the gauge transformation $\psi ^{A}\rightarrow \psi ^{A}-\partial
^{A}\Lambda $ , $\ $and $f$ is an arbitrary scalar function.

It is worthwhile to note that (\ref{p14}) is exactly the Faddeev-Popov
technique \cite{Faddeev} used to quantize a local gauge theory.
Consequently, the equation (\ref{p14}) can be expressed as being
\begin{equation}
Z=N\left( \beta \right) \!\!\!\!\!\int\limits_{periodic}\!\!\!\!\!\mathcal{D}%
\psi \det \left\vert \frac{1}{\sqrt{\zeta }}\square \right\vert \exp \left\{
\int_{\beta }d^{4}x\left[ \frac{1}{2}\psi ^{T}\eta ^{0}\left( i\beta
^{A}\partial _{A}-\gamma \right) \psi -\frac{1}{2\zeta }\left( \partial
_{A}\psi ^{A}\right) ^{2}\right] \right\} ~\ \ .  \label{pa16}
\end{equation}%
where the index $A=\tau ,1,2,3$ and $\beta ^{A}\partial _{A}=i\beta
^{0}\partial _{\tau }+\beta ^{k}\partial _{k}$. We remark that at
zero temperature the $\det \left( \square \right) $ is a constant
that can be ignored, however, at Finite Temperature it turns out a
very important temperature dependent term. Using the projectors
$R^{A}$ we rewrite the gauge fixing term in matrix form such that
the partition function reads as
\begin{equation}
Z=N\left( \beta \right) \det \left\vert \frac{1}{\sqrt{\zeta }}\square
\right\vert Z^{\prime }~\ \ ,  \label{pa-16}
\end{equation}%
where%
\begin{equation}
Z^{\prime }=\!\!\!\!\!\int\limits_{periodic}\!\!\!\!\!\mathcal{D}\psi \exp
\left\{ \int_{\beta }d^{4}x\left[ \frac{1}{2}\psi ^{T}\eta ^{0}\left( i\beta
^{A}\partial _{A}-\gamma +\frac{1}{\zeta }\eta ^{0}\left( R^{A}\right)
^{T}R^{B}\partial _{A}\partial _{B}\right) \psi \right] \right\} ~\ \ ,
\label{pa17}
\end{equation}
and $R^{A}\partial _{A}=iR^{0}\partial _{\tau }+R^{k}\partial _{k}$.

To perform the calculation of the functional integral we use the Fourier
series of the DKP field
\begin{equation}
\psi \left( \tau ,\mathbf{x}\right) =\frac{1}{\beta }\sum\limits_{n}\int
\frac{d^{3}\mathbf{p}}{\left( 2\pi \right) ^{3}}\widetilde{\psi }\left(
\omega _{n},\mathbf{p}\right) e^{i\left( \mathbf{px}+\omega _{n}\tau \right)
}~\ \ ,  \label{pa18}
\end{equation}%
with $\omega _{n}=2\pi n/\beta $ and the periodicity conditions (\ref{pa3})
for the DKP field are imposed.

Then, substituting (\ref{pa18}) in (\ref{pa17}) we get
\begin{eqnarray}
Z^{\prime } &=&\prod\limits_{n,\mathbf{p}}\det \left[ i\omega _{n}\beta
^{0}+p_{k}\beta ^{k}+\gamma +\frac{1}{\zeta }\eta ^{0}\left( i\omega
_{n}R^{0}+p_{k}R^{k}\right) ^{T}\left( i\omega _{n}R^{0}+p_{k}R^{k}\right) %
\right] ^{-1/2}  \notag \\
&=&\prod\limits_{n,\mathbf{p}}\left[ \frac{\left( \omega _{n}^{2}+\omega _{%
\mathbf{p}}^{2}\right) ^{4}}{\zeta }\right] ^{-1/2}=\prod\limits_{n,\mathbf{p%
}}\left( \omega _{n}^{2}+\omega _{\mathbf{p}}^{2}\right) ^{-2}\sqrt{\zeta }\
\ ~,  \label{det-19}
\end{eqnarray}%
where $\omega _{\mathbf{p}}=\left\vert {\mathbf{p}}\right\vert $. The
determinant in (\ref{pa-16}) is
\begin{equation}
\det \left\vert \frac{1}{\sqrt{\zeta }}\square \right\vert =\prod\limits_{n,%
\mathbf{p}}\frac{\left( \omega _{n}^{2}+\omega _{\mathbf{p}}^{2}\right) ^{1}%
}{\sqrt{\zeta }}~\ \ .  \label{det-20}
\end{equation}%
Finally, the partition function reads
\begin{equation}
Z=N\left( \beta \right) \prod\limits_{n,\mathbf{p}}\left( \omega
_{n}^{2}+\omega _{\mathbf{p}}^{2}\right) ^{-1}~\ \ ,
\end{equation}%
as we are interesting in $\ln Z$, then, from the last expression we obtain
\begin{equation}
\ln Z=\ln N\left( \beta \right) +2V\sum\limits_{n}\int \frac{d^{3}\mathbf{p}%
}{\left( 2\pi \right) ^{3}}\ln \beta -V\sum\limits_{n}\int \frac{d^{3}%
\mathbf{p}}{\left( 2\pi \right) ^{3}}\ln \beta ^{2}\left( \omega
_{n}^{2}+\omega _{\mathbf{p}}^{2}\right) ~\ \ .
\end{equation}%
The value for the $N\left( \beta \right) $ is selected in a manner that it
cancels the divergent term, i.e we choose
\begin{equation}
\ln N\left( \beta \right) =-2V\sum\limits_{n}\int \frac{d^{3}\mathbf{p}}{%
\left( 2\pi \right) ^{3}}\ln \beta ~\ \ ,
\end{equation}%
next, we perform the sum (see, for example, \cite{Bernard,libros}) and get
the following expression for the partition function
\begin{equation}
\ln Z=-2V\int \frac{d^{3}\mathbf{p}}{\left( 2\pi \right) ^{3}}\left[ \frac{%
\beta \omega _{\mathbf{p}}}{2}+\ln \left( 1-e^{-\beta \omega _{\mathbf{p}%
}}\right) \right] ~\ \ ,
\end{equation}%
which describe a massless bosonic field with two polarization states that is
the characteristic behavior of the electromagnetic field in thermodynamical
equilibrium.

\section{Conclusions}

In this work we study the massless DKP theory at Finite Temperature, and it
is shown the constraint structure of the model leads to conclude that it is
a local $U(1)$ gauge theory in its spin 1 sector. Such analysis allow to
construct the correct partition function using the Hamiltonian procedure.
Also, we show that it is possible to arrive to the covariant expression
which is exactly the covariant quantization of a gauge theory using the
Faddeev-Popov approach. Consequently, the partition function of the spin 1
sector gives the partition function of a zero-mass Bose gas with two
polarization states, i.e. the electromagnetic field modes in thermodynamical
equilibrium.

The perspectives to be followed are to study the Finite Temperature
properties of the massless DKP field in curved space-time and, consequently,
analyze the curvature effects in the thermodynamics of the electromagnetic
field via the DKP formalism. Advances in this directions will be reported
elsewhere.

\subsection*{Acknowledgements}

RC (grant 01/12611-7) thanks FAPESP for full support. BMP thanks CNPq and
FAPESP (grant 02/00222-9) for partial support.

\appendix

\section{Spin 1 representation for the massless DKP algebra}

\begin{center}
\scalebox{0.7}{\includegraphics{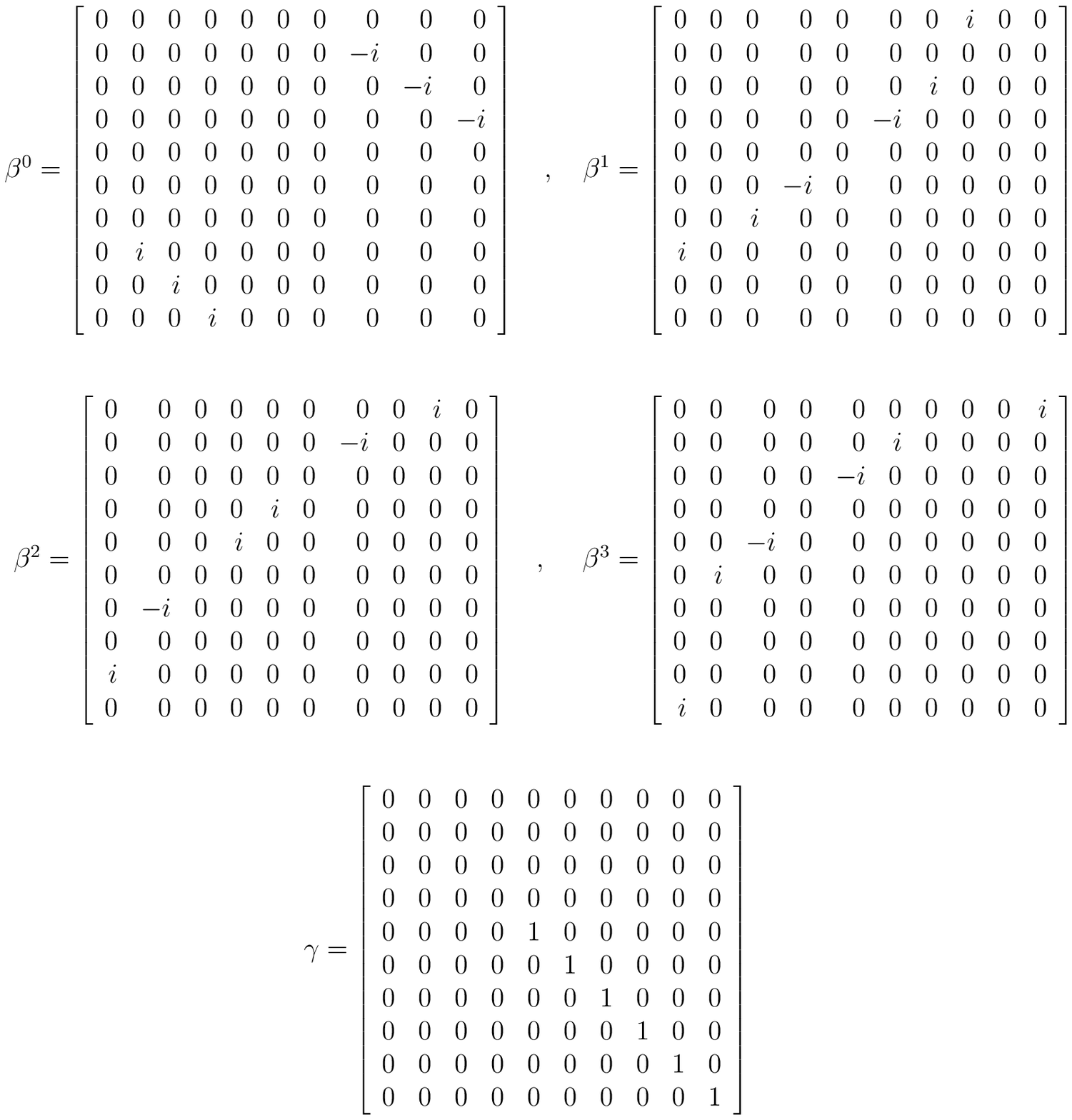}}
\end{center}

\end{document}